\documentclass[preprint,superscriptaddress,12pt]{elsarticle}
\usepackage{amssymb}
\usepackage{amsmath}
\usepackage{multibib}
\usepackage{multirow}
\usepackage{tabularx}
\usepackage{booktabs}
\usepackage{adjustbox}
\usepackage{threeparttable}
\usepackage{orcidlink}

\usepackage{color}
\usepackage{float}
\usepackage[a4paper, margin=2.5cm]{geometry} 
\usepackage{ulem}
\usepackage{hyperref}
\usepackage[labelfont=bf, labelsep=space]{caption}

\journal{Science Bulletin}

\title{A giant Peanut-shaped ultra-high-energy $\gamma$-ray emitter off the galactic plane}

\begin{document}

\begin{frontmatter}
\cortext[cor1]{Corresponding authors\\}
\author{LHAASO Collaboration\corref{cor1}\fnref{fn1}}
\ead{lizhe@ihep.ac.cn(Li Z), xishaoqiang@ihep.ac.cn(Xi SQ), chensz@ihep.ac.cn(Chen SZ)} 
\fntext[fn1]{Authors are listed at the end of this paper.}

\begin{abstract}
Ultra-high-energy (UHE) gamma($\gamma$) rays serve as unique probes for investigating new non-thermal astronomical phenomena within the Galaxy and also for studying PeV particle acceleration and propagation. Based on the UHE $\gamma$-ray survey conducted by LHAASO, this paper reports the discovery of a new source distinct from all previously identified  high energy $\gamma$-ray objects. This source exhibits an elongated, peanut-like structure, which can be described as a rectangular strip measuring 4.6$^\circ$ in length and 0.45$^\circ$ in width, featuring two bright knots. It is located at a Galactic latitude of $-$17.5$^\circ$, well below the Galactic plane and away from regions rich in UHE sources. The energy spectrum is spatially uniform across the entire region and is well fitted by a power-law model with an index of $-$1.95 and a cutoff at 250 TeV. The highest detected photon energy reaches 760 TeV. Morphological and spectral analyses indicate the presence of a powerful particle accelerator in this region, capable of accelerating particles beyond 1 PeV.  An energetic millisecond pulsar (MSP) J0218+4232 is located at the brighter end of this source, with a spatial chance coincidence probability of 3.6×10$^{-4}$, corresponding to 3.4$\sigma$. This suggests that the emission from this region very likely originates from particles accelerated within the MSP wind nebulae, potentially providing the first observational evidence supporting the wind nebulae of MSP as sources of UHE $\gamma$ rays. However, the presence of  other hidden accelerators, whether associated with known or entirely new populations of PeVatrons, cannot be ruled out, motivating future multi-wavelength observations.

\end{abstract}

\begin{keyword}

UHE $\gamma$-ray \sep Peanut \sep millisecond pulsar \sep anisotropic diffusion \\
Received: 22-Dec-2025\\
Revised: 26-May-2026\\
Accepted: 29-Aug-2026\\

\end{keyword}
\end{frontmatter}

\section{Introduction}
\label{sec:intro}
Ultra-high-energy (UHE, E$>$100 TeV) $\gamma$ rays occupy the highest energy end of the electromagnetic spectrum and serve as important probes for investigating particle acceleration and propagation processes at PeV energies in the non-thermal universe \cite{Cao:2023mig}. However, due to absorption by the cosmic microwave background (CMB), current observations of UHE $\gamma$ rays are limited to sources within our Galaxy. These $\gamma$ rays are primarily produced through inverse Compton scattering of high-energy electrons off CMB photons, or via collisions between high-energy hadrons and interstellar gas. The required particle energies and environmental conditions for these processes differ significantly from those governing radiation in the radio, X-ray, and even GeV–TeV $\gamma$-ray regimes. Consequently, UHE $\gamma$ rays provide a unique window for discovering new astronomical phenomena in the Galaxy.

In recent years, a new generation of ground-based extensive air shower (EAS) arrays, represented by the Large High Altitude Air Shower Observatory (LHAASO), has ushered in a new era in UHE $\gamma$-ray astronomy with its significantly improved sensitivity \cite{LHAASO:2021nature}. LHAASO’s survey observations have revealed more than 40~UHE sources, demonstrating the widespread presence of natural particle accelerators in the Galaxy capable of reaching PeV energies, and greatly advancing our understanding of the acceleration limits of Galactic sources \cite{LHAASO:2023rpg}. For example, LHAASO has confirmed pulsar wind nebulae (PWN) as dominant UHE $\gamma$-ray emitters \cite{LHAASO:2023rpg}, capable of accelerating electrons to multi-PeV energies with high efficiency \cite{LHAASO:2021cbz}; the discovery of a giant UHE $\gamma$-ray bubble in the Cygnus region suggests the presence of an unknown super-accelerator with energies exceeding 10~PeV \cite{LHAASO:2023uhj}; systematic observations of micro-quasars have reinforced their role as important PeV particle accelerators \cite{LHAASO:2024psv}; and both supernova remnants and compact binary systems have also been confirmed as UHE $\gamma$-ray sources \cite{LHAASO:2024fym,LHAASO:2025bwq}. Most of these identified UHE sources are distributed along the Galactic plane, and their spatial morphology—typically appearing as point-like or extended structures approximating a two-dimensional Gaussian profile—resembles that of sources detected in the GeV and TeV ranges.

PWNs are radiative regions formed by the interaction between relativistic winds driven by rapidly rotating neutron stars and the surrounding medium. At the termination shock of these winds, charged particles (primarily electrons and positrons) are efficiently accelerated. These particles then produce broad-band emission from radio to $\gamma$-ray wavelengths via mechanisms such as synchrotron radiation and inverse Compton scattering, with the spin-down energy of the pulsar supplying the power. Observations have confirmed that various evolutionary stages of PWN—including young PWN \cite{LHAASO:2023rpg,LHAASO:2021cbz,LHAASO:2024tbw}, aged pulsar halos \cite{LHAASO:2021crt}, and the tails of bow-shock pulsars \cite{LHAASO:2025rgh} can produce UHE $\gamma$-ray emission, underscoring their powerful particle acceleration capabilities and the rich diversity of their underlying physical processes. Millisecond pulsars (MSPs), characterized by spin periods less than 30~ms, typically reside in binary systems where they have been spun up via accretion from a companion star. Their relativistic winds may form structures similar to those of ordinary PWN and thus also produce $\gamma$-ray emission. However, observational evidence for such emission has remained elusive \cite{Hooper:2021kyp}.

The LHAASO-KM2A array, a main sub-array of LHAASO, offers a large field of view, continuous all-weather operation, and high sensitivity. Since early 2020, it has maintained a nearly 100\% operational duty cycle, covering approximately two-thirds of the sky each day, thereby providing a crucial opportunity to discover new phenomena in the UHE regime. In this paper, we report the discovery of a unique UHE $\gamma$-ray source, based on LHAASO survey data, located outside the Galactic plane. This source exhibits a distinctive peanut-shaped morphology that differs markedly from all previously known high-energy $\gamma$-ray sources. It shows a significant spatial coincidence with a high spin-down luminosity MSP J0218+4232, suggesting that its emission likely originates from a PWN associated with this MSP. This finding provides new observational clues for understanding particle acceleration and propagation in such systems.

\section{Observation}
\label{sec:result}
\subsection{Experiment  data}
\label{subsec:obser}
KM2A, the largest sub-array of LHAASO, covers an area of 1.36 km$^2$ and comprises 5216 electromagnetic detectors (EDs) and 1188 muon detectors (MDs). The ED data are used to reconstruct the energy and arrival direction of primary particles, achieving an energy resolution better than 20\% and an angular resolution of $0.25^\circ$ at 100 TeV \cite{Aharonian:2020iou}. The number of muons recorded by the MDs enables the rejection of more than 99.99\% of the cosmic-ray background above 100 TeV \cite{Aharonian:2020iou}, giving KM2A the world's highest sensitivity to UHE $\gamma$ rays.

This study uses KM2A data collected from 27 December 2019 to 31 January 2024, with a total live time of 1384~days. We apply the standard KM2A data analysis pipeline as described in \cite{LHAASO:2023rpg,Aharonian:2020iou}. Detector performance has been extensively studied using Monte Carlo (MC) simulations \cite{LHAASO:2024rtc}, and has been calibrated and monitored the Crab nebula as a standard candle \cite{Aharonian:2020iou, LHAASO:2024kbg}. In the standard analysis, all $\gamma$-like events in the energy range from 10 TeV to 10 PeV are grouped into 12 logarithmic energy bins and mapped into sky maps with a spatial pixel size of $0.1^\circ \times 0.1^\circ$ in celestial coordinates. The background, dominated by residual cosmic rays, is estimated using the direct integral method \cite{Milagro:2003yym}. The region of interest (ROI) used in this study is a $10^\circ \times 10^\circ$ region centered at right ascension (R.A.) $\alpha_\mathrm{J2000}=34.6^\circ$ and declination (Decl.) $\delta_\mathrm{J2000}=42.9^\circ$.  Given the low expected $\gamma$-ray flux at high Galactic latitudes ($|b| > 10^\circ$), Galactic diffuse emission (GDE) background is neglected in this analysis.

\subsection{The morphology}
\label{subsec:mophology}
\begin{figure}[h!]
    \centering
    \includegraphics[height=0.62\textwidth,width=0.50\textwidth]{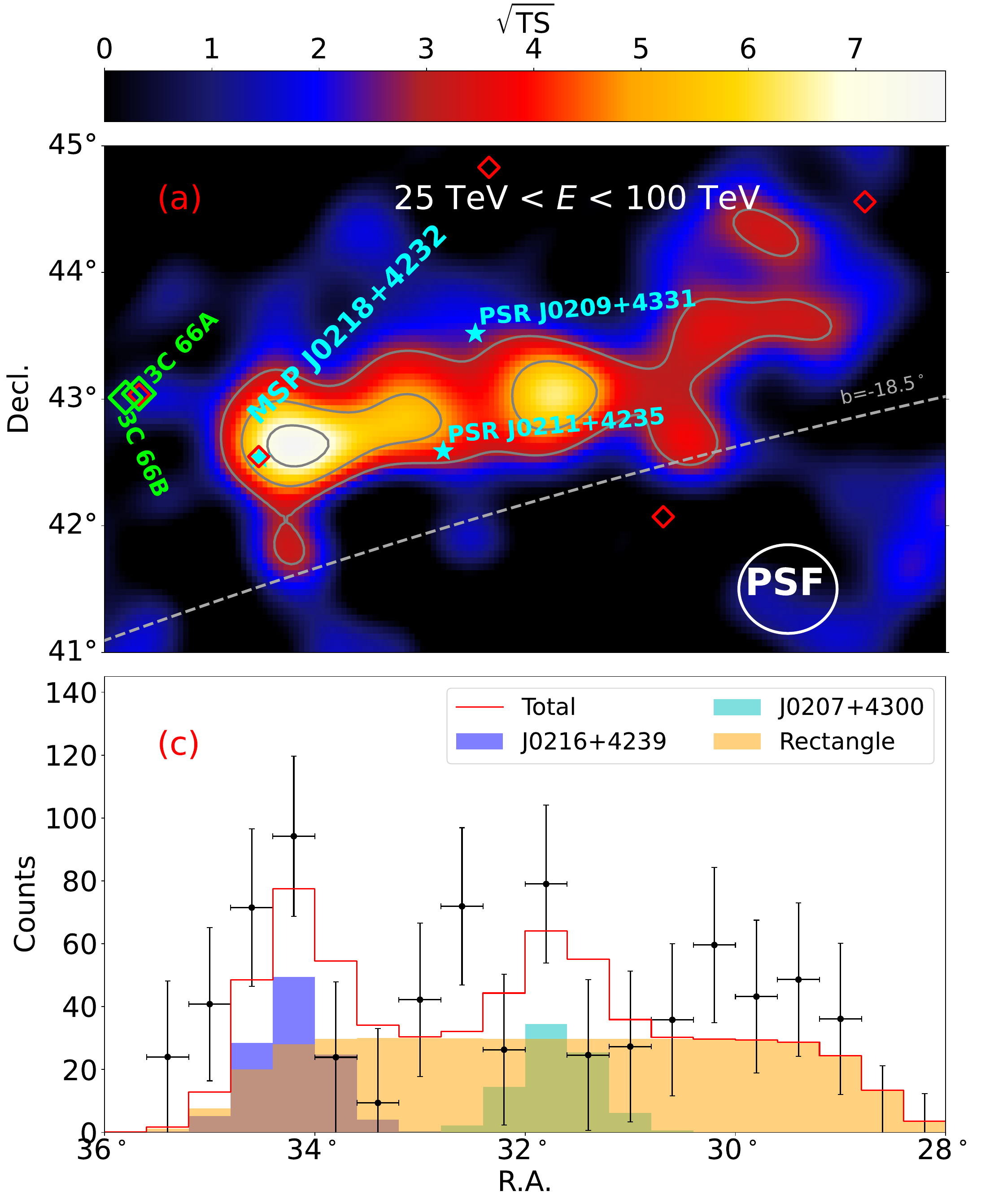}%
    \includegraphics[height=0.62\textwidth,width=0.50\textwidth]{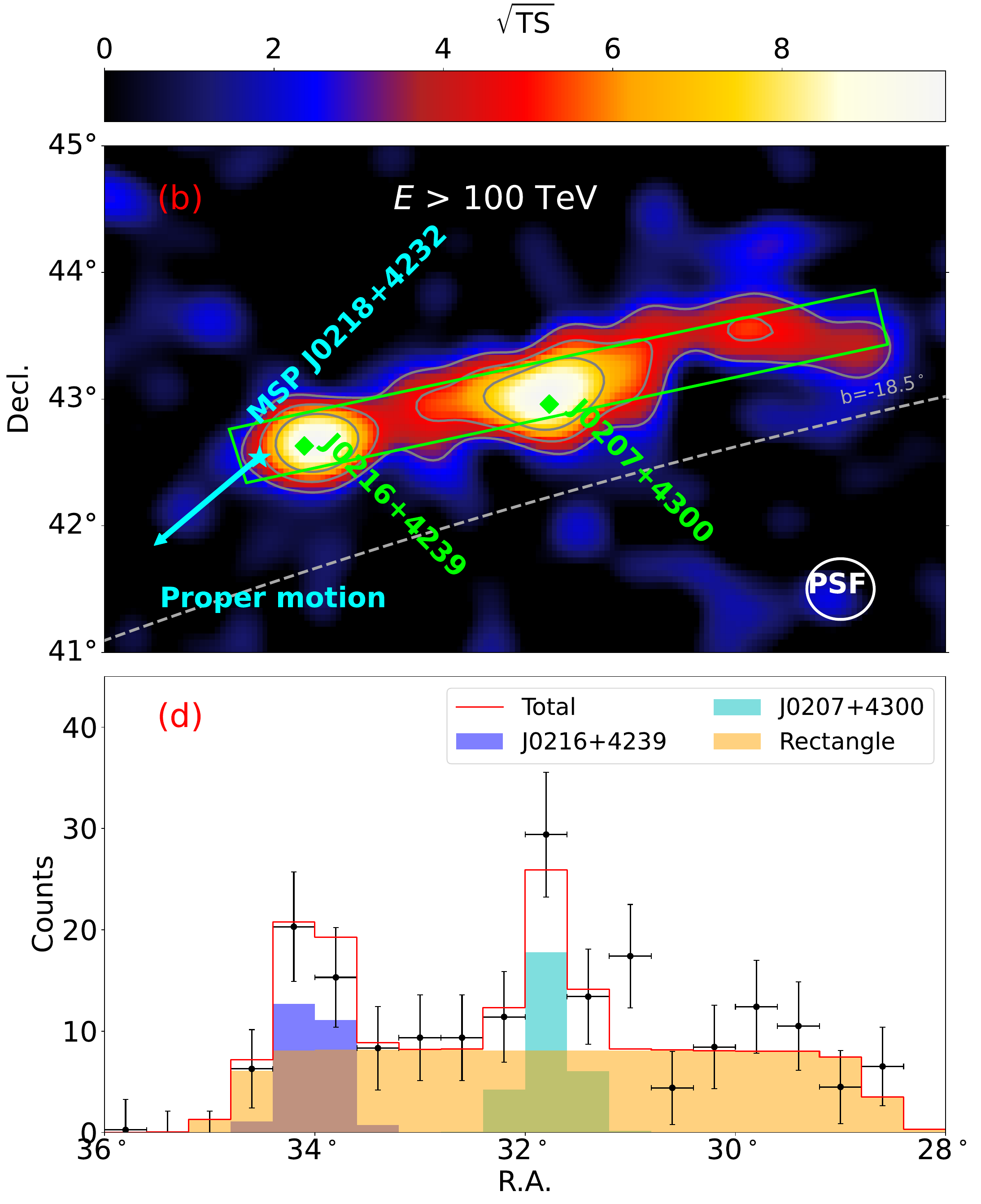}
    \caption{Significance maps  of the Peanut in two energy intervals,  25-100 TeV~(a) and above 100 TeV~(b).  The angular resolution (68\% containment) is indicated by the white circles. Contours represent the significance levels of 3$\sigma$, 5$\sigma$ and 7$\sigma$. The  green squares mark the TeV sources. The cyan stars are the pulsars. The red squares represent GeV sources. The rectangle shows the template for the strip-like diffuse emission component. The $\gamma$-ray event profiles in the R.A. direction are shown in (c) and (d) for two energy bands, respectively. The error bar represents a statistical uncertainty at the 68\% c.l. } 
    \label{fig:skymap}
\end{figure}

To reach better angular resolution, only events with energies above 25 TeV were used for morphological analysis. Significance maps in the 25$-$100 TeV and >100 TeV bands centered on the target source are shown in Fig.~\ref{fig:skymap}. These maps were produced using the standard pipeline described in \cite{LHAASO:2023rpg}, which moves a putative point source across the sky pixels and performs a likelihood-ratio test between background-only and single-source models. The source spectrum was assumed to follow a power law with an index of $-$2.7. According to Wilks’ theorem for one degree of freedom (dof), the square root of the test statistic~(TS) value corresponds to the detection significance \cite{Wilks:1938dza}. A clear $\gamma$-ray excess resembling a ``peanut'' shape—henceforth denoted as the Peanut—is visible in both maps, with a maximum significance exceeding $7\sigma$. The Peanut lies well outside the Galactic Plane, spanning Galactic longitudes between $135^\circ$ and $140^\circ$ at a latitude of $b \approx -17.5^\circ$.

To accurately characterize the spatial morphology of the Peanut, we performed a forward-fitting analysis on the ROI using a binned 3-dimensional likelihood method, testing different geometric models. For simplicity, all spectra were modeled with the power-law function. The baseline analysis followed the approach in \cite{LHAASO:2023rpg}, iteratively adding Gaussian components. A new component was considered statistically significant if it improved the fit by $\Delta\text{TS} \geq 25$ (corresponding to $\sim3.8\sigma$ for dof=5). This procedure identified four significant components, as listed in Table~\ref{tab:fourcomponents}.

These four components exhibit distinctive characteristics, which we examine in terms of their spectral properties and spatial distribution. LHAASO's survey at energies above 25 TeV \cite{LHAASO:2023rpg} showed that $\gamma$-ray sources are predominantly concentrated along the Galactic plane, with very few detected at high Galactic latitudes. Given a source number density of approximately 0.00126 deg$^{-2}$ in this region, the chance probability of four unrelated $\gamma$-ray sources clustering within the Peanut area is $2.7\times10^{-9}$ ($5.8\sigma$). Moreover, whereas most $\gamma$-ray sources exhibit spectral indices between $-$3 and $-$4, all four components here show exceptionally hard and similar spectra, with indices around $-$2.6. Furthermore, their spatial distribution reveals a striking alignment. Taken together, these distinctive features strongly suggest a common physical origin for these four components.

\begin{table}[h!]
 \centering
 \footnotesize 
 \setlength{\tabcolsep}{0.03in}
 \caption{Multi-Gaussian component analysis results. }

 \begin{tabular}{lcccccccccc}
 \hline
   & $J_0\times 10^{-17}$                                              & $\Gamma$ & $\alpha_\mathrm{J2000}$ & $\delta_\mathrm{J2000}$ &$l$    & $b$    &$p_\mathrm{95}$ & $r_\mathrm{39}$ & TS & TS$_\mathrm{ext}$ \\
     & ($\rm\ cm^{-2}\ s^{-1}\ TeV^{-1}$) &          & ($^\circ$)               & ($^\circ$)  & ($^\circ$) & ($^\circ$) & ($^\circ$)    & ($^\circ$)      &    & \\
\hline
J0216+4238 & 1.77$\pm$0.21 & $-$2.68$\pm$0.15 & 34.08 & 42.64 & 139.15 & $-$17.55 & 0.06 & $<$ 0.11 & 58.1 & 0.1\\
J0211+4255 & 1.64$\pm$0.26 & $-$2.57$\pm$0.17 & 32.87 & 42.91 & 138.17 & $-$17.58 & 0.14 & 0.26$\pm$0.05 & 61.5 & 16.1\\
J0207+4300 & 1.51$\pm$0.21 & $-$2.66$\pm$0.16 & 31.72 & 43.04 & 137.29 & $-$17.72 & 0.09 & $<$ 0.19 & 36.7 & 4.9\\
J0201+4330 & 3.00$\pm$0.41 & $-$2.60$\pm$0.14 & 30.07 & 43.50 & 135.94 & $-$17.63 & 0.27 & 0.55$\pm$0.06 & 74.7 & 35.6\\
\hline
 \end{tabular}   
\label{tab:fourcomponents}

\begin{tablenotes}
   \footnotesize
  \item{Notes: The spectrum was assumed to be a power-law (PL) $\rm{d}N/\rm{d}E=J_0(E/E_0)^{\Gamma}$, where $J_0$ is the normalization, $E_0$ is fixed to 50 TeV, and $\Gamma$ denotes the index; $\alpha_{J2000}$ and $\delta_{J2000}$ are the right ascension and declination with standard equinox J2000, respectively. $l$ and $b$ are the Galactic longitude and Galactic latitude, respectively. $p_{95}$ is the error of the center position at the 95\% confidence level (c.l.). $r_{39}$ is the extension size corresponding to 39\% of the source flux. The TS is for each component. TS$_{\rm ext}$ is obtained from a test between the extended-source and point-source hypotheses.}  
   \end{tablenotes}

\end{table}

Given the morphological coherence and spectral consistency, we further tested the presence of a diffuse emission component using a 5-parameter rectangular template. Starting from this strip-like diffuse model, we added Gaussian or point sources from Table~\ref{tab:fourcomponents} and performed a joint fit of all spatial and spectral parameters. Including the strip component significantly improved the fit over the four-Gaussian model, with $\Delta\text{TS} = 37.8$ for dof$=$7, corresponding to a detection significance of $4.5\sigma$. Within this combined model, two components remained significant ($>4\sigma$), while the other two became negligible ($<3\sigma$). A summary of model comparisons is provided in Table~\ref{tab:mor}. For reference, the AIC values are also listed in the table. The final favored morphology of the Peanut is thus a composite of a strip-like structure and two point-like knots, as marked in Fig.~\ref{fig:skymap}.

The strip is oriented $13.8^\circ \pm 1.2^\circ$ north of west, with a length of $4.60^\circ \pm 0.44^\circ$ and a width of $0.45^\circ \pm 0.12^\circ$, roughly parallel to the Galactic plane (see Fig.~\ref{fig:skymap}). The Peanut is more significant at energies above 100~TeV, where the overall significance reaches $16.0\sigma$, compared to $9.7\sigma$ in the 25$-$100~TeV band. Projected $\gamma$-ray event profiles along the right ascension clearly show two peaks superposed on the strip, as illustrated in Fig.~\ref{fig:skymap}. The energy dependence of the Peanut's morphology was investigated, but no significant variation was found (see Section S1 in the Supplementary material).

\begin{table}[h!]
   \centering
    \caption{Morphological analysis results for the Peanut region.}
\small
    \setlength{\tabcolsep}{5mm}{
    \begin{tabular}{lccccc}
    \hline
     & Spatial Model  & TS   & dof& AIC \\
\hline
MD1 (baseline) & Four components & 430.4 & 18 & 0\\

MD2 & Four components + strip & 468.2 & 25 & $-$23.8 \\

MD3 & Three components + strip & 463.6 & 20 & $-$29.2 \\

MD4 (favored) & Two components + strip & 446.8 & 15 & $-$22.4 \\

MD5 & One component +strip & 417.8 &11 & $-$1.4 \\
\hline
   \end{tabular}}
   \begin{tablenotes}
   \footnotesize
  \item{Notes: MD1 is shown in Table \ref{tab:fourcomponents}. The strip is modelled with a rectangle. TS values are obtained by likelihood estimation with a background-only model and corresponding spatial models listed in this table. AIC is a statistical quantity that measures the goodness of fit of different models, defined by AIC = 2$k$$-$2ln$\mathcal{L}$, where k is the number of parameters in the model. In this formulation, the best hypothesis is considered to be the one that minimizes the AIC. The AIC value of MD1 is subtracted from that of each model for a clear comparison.}  
   \end{tablenotes}
   \label{tab:mor}
\end{table}

\begin{figure}[h!]
    \centering
     \includegraphics[width=0.85\textwidth]{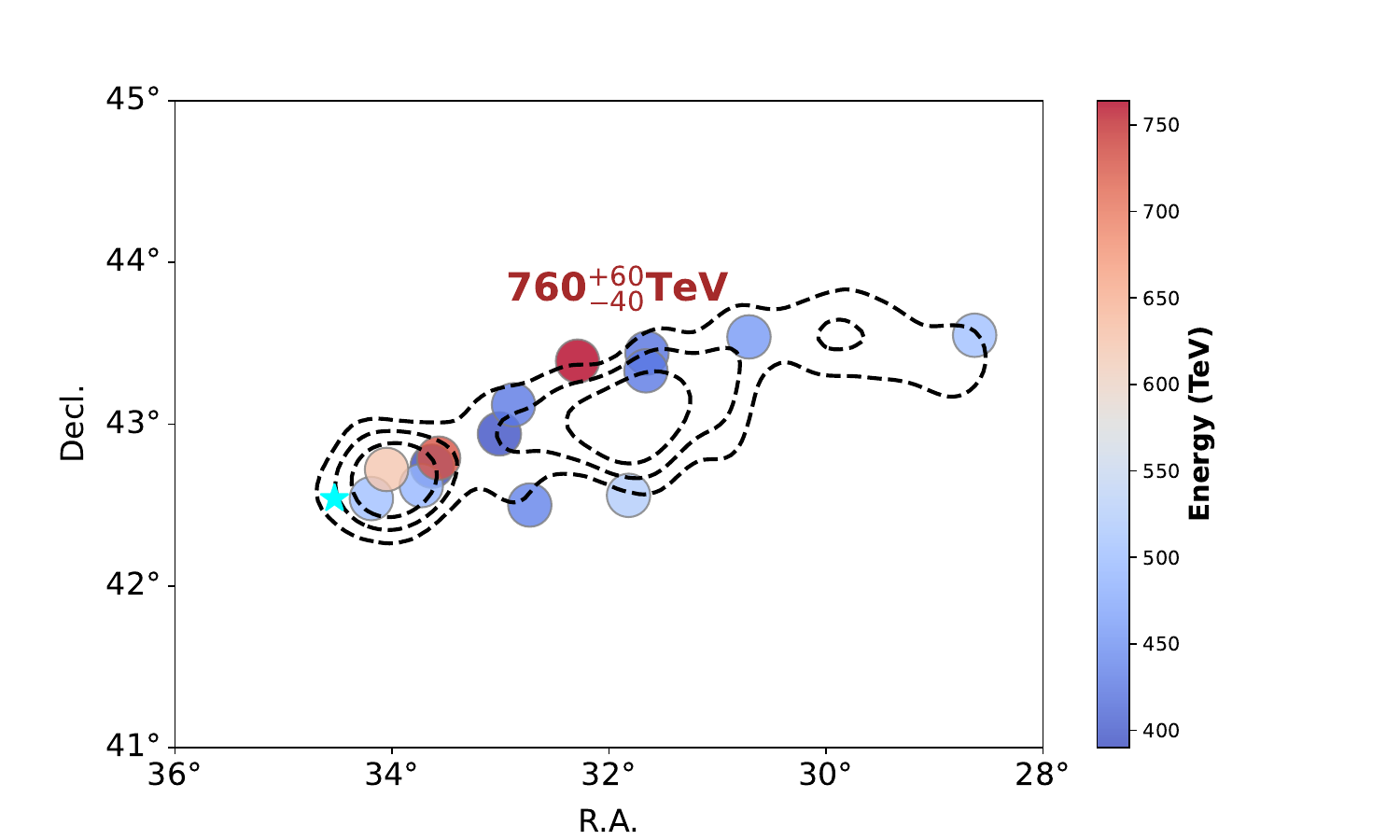}
    \caption{Fourteen photons at energies above 400~TeV were detected by LHAASO from the Peanut region. The number of background events, dominated by residual cosmic rays, in this region is estimated to be 3.2. The dashed contours indicate the statistical significance of photons above 100 TeV. Additionally, the cyan star indicates the position of the MSP J0218+4232.}
    \label{fig:event} 
\end{figure}

Notably, 14 photons with energies exceeding 400 TeV have been detected within the Peanut region, compared to an expected background of 3.2 events, corresponding to a  4.4$\sigma$ $\gamma$-ray excess. Their spatial positions are shown in Fig.~\ref{fig:event}. Five of these photons are in the first knot, while the remaining nine are distributed across other parts of the structure, preferentially near its expected boundary. To accurately estimate the highest photon energies, we re-evaluated their values and uncertainties using a Bayesian approach, incorporating the best-fit spectrum from the following section and individual photon characteristics, as detailed in our previous work \cite{LHAASA:2023pay}. The six highest-energy photons are listed in Section S3 in the Supplementary material, with the most energetic event reaching $760_{-40}^{+60}$ TeV, located at the edge of the second knot.

\subsection{The spectrum}
\label{subsec:spectr}
Following the morphological characterization, we performed a detailed spectral analysis of the $\gamma$-ray emission from the Peanut region. The analysis also employed the 3D maximum-likelihood fitting method and covered an energy range from 10~TeV to 10~PeV, which was logarithmically divided into 12 bins. Given the spectral similarity among the different components, a single spectral shape was adopted for all three. A power law with an exponential cutoff (PLEC) was found to improve the fit by $\Delta \text{TS} = 17$ relative to a simple power law, corresponding to a significance of $4.1\sigma$. The flux points are listed in Section S2 in the Supplementary material, and the PLEC spectrum is
\begin{equation} 
\mathrm{d}N/\mathrm{d}E=(8.10\pm1.0)\times 10^{-17}~(E_{\gamma}/50 {\rm\ TeV})^{-(1.95\pm0.19)}\mathrm{e}^{-E/(250\pm82{\rm\ TeV})}\rm\ cm^{-2}\ s^{-1}\ TeV^{-1}.
\end{equation}

\begin{figure}[h!]
    \centering
     \includegraphics[width=0.85\textwidth]{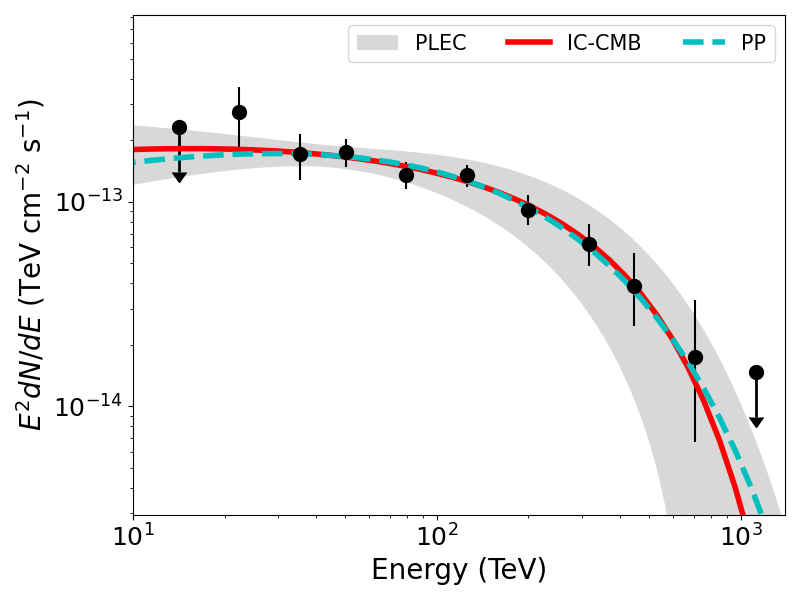}
    \caption{Spectral energy distribution of the entire Peanut above 10 TeV observed by LHAASO. Flux points are marked with black dots. The error bar represents a statistical uncertainty at the 68\% c.l., two flux upper limits are at 95\% c.l. }\label{fig:spectrum} 
\end{figure}

As shown in Fig.~\ref{fig:spectrum}, the gray band shows the $\pm1\sigma$ uncertainty range of the fit. Table~\ref{tab:peanutthree} summarizes the best-fit parameters for these three components. The strip-like diffuse component dominates the emission, accounting for 72\% of the total flux from the Peanut region, while the other two components each contribute approximately 14\%.

\begin{table}[h!]
 \centering
 \footnotesize 
 \setlength{\tabcolsep}{0.03in}
 \caption{Spectral and morphology parameters of all Peanut components. }
 \begin{tabular}{lcccccccccc}
 
 \hline
 & $J_0\times 10^{-17}$ & $\Gamma$ & $E_\mathrm{cut}$ & $\alpha_\mathrm{J2000}$ & $\delta_\mathrm{J2000}$  & $p_\mathrm{95}$ & Length & Width & Orientation  \\
 & $(\rm\ cm^{-2}\ s^{-1}\ TeV^{-1})$ &          & (TeV) & ($^\circ$)               & ($^\circ$)   & ($^\circ$) & ($^\circ$) & ($^\circ$)     &  angle($^\circ$)      &  \\
\hline
The strip  & 5.82$\pm$0.56 & $-$1.95$\pm$0.19 & 250$\pm$82& 31.70 & 43.14 & 0.11 & 4.60$\pm$0.44 & 0.45$\pm$0.12 & 13.8$\pm$1.2\\
J0216+4239 & 1.05$\pm$0.23 & - & - & 34.10 & 42.63 & 0.09 & - & - & -  \\
J0207+4300 & 1.20$\pm$0.24 & - & - & 31.77 & 42.96 & 0.12 & - & - & -   \\
\hline
 \end{tabular}  
    \begin{tablenotes}
   \footnotesize
  \item{Notes: $J_0$ is the normalized flux at 50 TeV, $\Gamma$ is the power-law index, and $E_\mathrm{cut}$ is the exponential cutoff energy. $\alpha_\mathrm{J2000}$ and $\delta_\mathrm{J2000}$ denote the right ascension and declination in the J2000 equatorial coordinate system, respectively, and $p_{95}$ is the 95\% confidence-level positional uncertainty of the source center. The length, width, and position angle relative to the R.A. axis of the strip are also listed. The estimated systematic uncertainties are less than $0.01^\circ$ for the position, approximately $0.08^\circ$ and $0.008^\circ$ for the strip width and length, respectively, $\sim10\%$ for the flux, and $\sim0.03$ for the spectral index.}  
   \end{tablenotes}
 
\label{tab:peanutthree}
\end{table}

\subsection{Systematic uncertainties}\label{Sec:systematic}
 To assess the robustness of this observation result, we performed a number of systematic checks similar to those employed in Ref.~\cite{LHAASO:2023rpg}. The pointing error was estimated to be less than $0.01^\circ$ from observations of the standard candle, the Crab Nebula, as well as the cosmic-ray Moon shadow \cite{LHAASO:2024kbg}. For the extension estimation, the dominant contribution comes from the difference in the point spread function (PSF) between the MC simulation and the data. This difference was estimated to be about $0.08^\circ$ using observations of the point-like source Crab Nebula \cite{LHAASO:2023rpg}. Taking this difference into account, the systematic error in the strip extension measurement was estimated to be $0.008^\circ$ in length and $0.08^\circ$ in width.
 
The systematic uncertainties in the flux are primarily attributed to the atmospheric model used in the MC simulations. Due to seasonal and daily variations, the actual atmospheric density profile continuously deviates from the model. To evaluate this effect, the data were divided into four subsets according to daily atmospheric pressure: $<$594 hPa, 594$-$597 hPa, 597$-$600 hPa, and $>$600 hPa. The maximum difference observed among these subsets was adopted as a conservative estimate of the systematic error, resulting in a flux uncertainty of 10\% and a spectral index uncertainty of 0.03. When adopting a PLEC spectral model, the systematic uncertainty in the spectral index propagates to the cutoff energy ($E_{\mathrm{c}}$). By varying the spectral index within its uncertainty range to fit the Peanut spectrum, we obtained a variation of 8\% for $E_{\mathrm{c}}$, which we assigned as the conservative systematic uncertainty for this parameter. Additionally, uncertainties related to GDE, originating from cosmic‑ray interactions with interstellar gas, were assessed by including a dust template in the spectral fitting. The influence on both the extension and the total flux was found to be below 2\%.

\subsection{Parent particles responsible for the Peanut emission}
High-energy $\gamma$ rays cannot be produced through direct acceleration; their detection in the Peanut region therefore implies that energetic particles are either accelerated within or propagated into this area before radiating. The emission mechanisms for such $\gamma$ rays are broadly categorized into leptonic and hadronic processes.

In the leptonic scenario, $\gamma$ rays are primarily generated via inverse Compton scattering, where high-energy electrons upscatter low-energy photons. Due to the Klein–Nishina effect, CMB photons dominate as the seed photons for producing UHE $\gamma$ rays. Given the isotropy of the CMB, the properties of the underlying high-energy electron population can be directly inferred from the observed $\gamma$-ray spectrum of the Peanut. 

In the hadronic scenario, $\gamma$ rays arise mainly from collisions between high-energy hadrons—typically assumed to be protons—and the interstellar medium (ISM). Multi-wavelength observations toward the Peanut direction show no significant molecular clouds but rather an approximately uniform atomic hydrogen cloud (see Section S4 in the Supplementary material). Once the ISM density is constrained, the properties of the responsible high-energy proton population can be estimated.

Based on the shape of the observed $\gamma$-ray spectrum, we model the energy distribution of the radiating relativistic particles with a PLEC:
\begin{equation}\label{splc}
    \frac{\mathrm{d}N_{\rm p,e}}{\mathrm{d}E_{\rm p,e}}\propto E_{\rm p,e}^{\alpha} \exp\left[-\left(\frac{E_{\rm p,e}}{E_\mathrm{c}}\right)^\beta\right].
\end{equation}
where $E_{\rm p,e}$ is the primary particle energy, $\alpha$ is the spectral index, $E_{c}$ is the cutoff energy, and $\beta$ controls the sharpness of the cutoff. For electrons, we adopt $\beta=2$ to account for the pronounced effects of escape and radiative cooling. For protons, radiative cooling is negligible given their much longer cooling timescale; we therefore use  $\beta=1$ in the fitting. 

The $\gamma$-ray fluxes predicted by two scenarios are shown in Fig.~\ref{fig:spectrum}. The cyan dashed line represents the spectrum from proton–proton~(pp) interactions, corresponding to a proton spectrum of $(E_\mathrm{p}/{1\ \rm TeV})^{-1.45\pm0.55} \exp\left[-(E_\mathrm{p}/(0.86\pm0.48\ \rm PeV))\right]$. 
Assuming a target density of \(1\)~cm\({}^{-3}\), the total energy in protons above 100 TeV is  $W_\mathrm{p} = (3.6\pm0.3) \times 10^{48}\ \times (\frac{d}{3.15 \rm kpc})^{2} \ \rm erg$.
The red solid line corresponds to the inverse Compton spectrum from CMB photons scattered by UHE electrons, with an electron spectrum of $(E_\mathrm{e}/{1\ \rm TeV})^{-2.48\pm0.27} \exp\left[-(E_\mathrm{e}/(0.77\pm0.24\ \rm PeV))^2\right]$. The total energy in electrons above 50 TeV is $W_\mathrm{e} = (1.2\pm0.3) \times 10^{45}(\frac{d}{3.15 \rm kpc})^{2} \ \rm erg$.

\section{Interpreting the Peanut emission}
\subsection{The potential accelerator: MSP J0218+4232}
\label{Sec:multiwave}

UHE $\gamma$-ray sources in our Galaxy are typically associated with extreme astrophysical accelerators, such as PWN, SNRs, young massive star clusters (YMSCs), microquasars, and compact binary systems. To identify possible counterparts for the Peanut, we cross-matched its position with multiple catalogs, including those of SNRs/PWNe \cite{2012AdSpR..49.1313F}, the Fourth Fermi-LAT Catalog \cite{Ballet:2023qzs}, TeVCat \footnote{http://tevcat.uchicago.edu}, the ATNF pulsar catalog \footnote{http://www.pulsarastronomy.net/wiki/Main/PulsarCatalogues}, and X-ray binary catalogs. The resulting candidate counterparts are shown in the top-left panel of Fig.~\ref{fig:skymap}. The two nearby TeV sources, 3C66A and 3C66B, are extragalactic and thus unlikely to be physically related. Two pulsars, PSR J0211+4235 (distance $d=2.2$ kpc, spin-down luminosity $\dot{E}=1.0\times10^{30}$ erg s$^{-1}$) and PSR J0209+4331 ($d=2.5$ kpc, $\dot{E}=3.5\times10^{30}$ erg s$^{-1}$), are spatially adjacent but possess insufficient spin-down power to account for the observed $\gamma$-ray emission. An energetic MSP, J0218+4232 ($d=3.15$ kpc, $\dot{E}=2.4\times10^{35}$ erg s$^{-1}$), emerges as the only plausible candidate accelerator for the Peanut. 

MSP J0218+4232 is located at the bright head of the Peanut structure. Using the estimation method described in our previous work \cite{LHAASO:2023rpg}—which accounts for the pulsar’s spin-down luminosity and the local number density of pulsars—we find that the chance coincidence probability of a pulsar with a spin-down luminosity equal to or greater than that of MSP J0218+4232 appearing within a $0.5^{\circ}\times0.5^{\circ}$ region at the head of the Peanut is $3.6\times10^{-4}$. This corresponds to with a significance of $3.4\sigma$.

UHE $\gamma$-ray emission has been observed from numerous young and middle-aged pulsars, with the $\gamma$-ray luminosity scaling approximately with the spin-down power.
The spin-down luminosity of MSP J0218+4232 ranks fourth among all known MSPs within the LHAASO field of view. It becomes the brightest if we restrict the sample to MSPs at Galactic latitude $|b|>10^\circ$ with known distances. Therefore, this MSP is expected to be capable of powering $\gamma$-ray emission, and it is unsurprising that the first UHE $\gamma$-ray source associated with a MSP is detected from J0218+4232. Using the distance of MSP J0218+4232, the total $\gamma$-ray luminosity of the Peanut above 10 TeV is estimated to be $L_\mathrm{\gamma}=9.36\times10^{32}$ erg s$^{-1}$. The spin-down power of MSP J0218+4232 can supply this energy if the conversion efficiency ($\kappa_{\mathrm{UHE}}$) of rotational energy into UHE electrons or protons exceeds $4\times10^{-3}$. 

Adopting a leptonic halo or hadronic bubble scenario similar to that described in \cite{LHAASO:2023uhj,LHAASO:2021crt,2017Sci...358..911A}, the total energy in radiating electrons or protons required to explain the $\gamma$-ray flux detected by LHAASO is $W_\mathrm{e} \sim 1.2\times10^{45}$ erg or $W_\mathrm{p} \sim 3.6\times10^{48}$ erg (see Fig.~\ref{fig:spectrum}). For a fiducial conversion efficiency of $\kappa_{\mathrm{UHE}}=10^{-2}$, the time ($\tau \approx \frac{W}{\kappa_{\mathrm{UHE}} \dot{E}}$) required for MSP J0218+4232 to inject this population of electrons or protons is 
$16\ \text{kyr}$ or $50\ \text{Myr}$. The proper motion of the pulsar, measured as \(6.5\)~mas yr\({}^{-1}\) and oriented roughly parallel to the major axis of the strip component \cite{Du:2014dva}. 
In the leptonic scenario,  the angular displacement of the pulsar due to its proper motion is about 2 arcminutes over the characteristic formation timescale of 16 kyr. Considering the propagation of electrons, it is physically reasonable to explain the observed extension of the Peanut. In the hadronic scenario,  the angular displacement of the pulsar due to its proper motion is about 90$^\circ$ over the characteristic formation timescale of 50 Myr.   This displacement greatly exceeds the actual angular extension. Additionally, we investigated the gas distribution around the MSP and detected no morphological similarities~(see Section S4 in the Supplementary material). Given these discrepancies, the scenario whereby MSP J0218+4232 supplies all protons responsible for the entire $\gamma$-ray emission of the Peanut faces significant challenges and will not be further discussed herein.

\subsection{Possible origins of the Peanut's spatial morphology}

The spatial association between MSP J0218+4232 and the Peanut differs markedly from that observed for most known PWN and TeV halos. Typically, the pulsar lies within or near the center of its associated PWN or TeV halo. This common morphology is attributed to the fact that most such systems reside in the Galactic plane, where accelerated electrons are largely confined within the nebula or undergo slow diffusion in the surrounding medium after escaping the acceleration region. In contrast, MSP J0218+4232 is located at one end of the Peanut structure, and no significant extended emission is detected immediately around the pulsar itself.

A recently detected UHE source associated with the pulsar PSR J1740+1000 at high Galactic latitude \cite{LHAASO:2025rgh} presents a potentially analogous case. Similar to the Peanut, this source shows no evidence of the slow diffusion expected for pulsars of comparable age in the Galactic plane, and its UHE emission is significantly offset from the pulsar, located in the region opposite to its proper motion direction. MSP J0218+4232 also lies at high Galactic latitude $b \approx -17.5^\circ$. As indicated in Fig.~\ref{fig:skymap}, its proper motion velocity is 98~km s$^{-1}$ \cite{2014Verbiest}. Notably, the Peanut is situated roughly in the region opposite to this proper motion direction, mirroring the case of PSR J1740+1000, albeit with a much larger angular extent.

A plausible scenario is that electrons are accelerated at the termination shock, propagate mainly tailward, and then escape rapidly along the ordered Galactic magnetic field.
This interpretation is supported by the fact that the Peanut's major axis is approximately aligned with the local large-scale Galactic magnetic field direction. Recent simulations tracing the escape of energetic particles from accelerator sites demonstrate that in regions dominated by a regular large-scale Galactic field, particles propagate primarily along magnetic field lines \cite{Bao:2024rrg}. Consequently, the resulting emission morphology traces the magnetic field direction, with enhanced brightness predicted at magnetic field bends. If this picture holds, the Peanut's spatial structure directly traces the local large-scale Galactic magnetic field, providing empirical support for current Galactic magnetic field models \cite{2012ApJ...757...14J}. Within this framework, the bright knot at the head of the Peanut likely marks the compact pulsar tail region. This knot hosts numerous photons above 400 TeV as shown in Fig.~\ref{fig:event}, consistent with a primary particle transport zone. The knot in the central part of the Peanut could correspond to a magnetic field bend occurring primarily in the plane of the sky.

Supporting evidence for this general picture comes from X-ray observations, which have revealed several instances of high-energy electrons accelerated near middle-age pulsars and subsequently channeled away along the local large-scale Galactic magnetic field, manifesting as an X-ray ``pulsar filament'' or a ``misaligned outflow'' \cite{Klingler:2016exg,Klingler:2020rwb,Klingler:2023upn,Pavan:2013lka}. The Peanut represents a notable extension of this phenomenon to vastly larger scales. At the assumed distance of 3.15 kpc, its 4.6$^{\circ}$ length corresponds to a physical scale of approximately 250 pc. Considering that the Peanut's orientation is offset by only about 20° from the direction opposite to the pulsar's proper motion, if the bulk particle flow is indeed tailward, then upon entering the large-scale Galactic field, the particles would primarily follow the field lines. This alignment would result in weak synchrotron cooling in the direction perpendicular to the field, leading to a long cooling timescale and thereby enabling the electrons to propagate to the considerable distances observed.

It is worth noting that although the spatial morphology of the Peanut structure can be explained by the above scenario, its exact origin remains an open question. 

\subsection{Diffusion coefficient estimation}
\label{Sec:cooling}
Particle diffusion is considered as the dominant process responsible for shaping the Peanut. The diffusion distance is defined as $r_\mathrm{d} = 2\sqrt{Dt}$, where $D$ is the diffusion coefficient along a given direction and $t$ is the effective injection time of the leptons.
To produce UHE $\gamma$ rays via inverse Compton scattering of CMB photons, the energy of the primary $e^{\pm}$ can be estimated as $E_\mathrm{e} = 2.15\times(E_{\gamma}/1\ {\rm PeV})^{0.77}\ {\rm PeV}$ \cite{LHAASO:2021cbz}. The $\gamma$-ray energy range detected by LHAASO thus corresponds to parent electron energies from approximately 50 TeV to 1.6 PeV. For electrons in this energy range, the radiative cooling time, considering both synchrotron and inverse Compton losses in a typical Galactic halo magnetic field of $1\ \mu{\rm G}$ \cite{2012ApJ...757...14J}, is $t_\mathrm{c} = 14.3\times(E_{e}/400\ {\rm TeV})^{-0.5}$ kyr (see Section S5 in the Supplementary material). Given that the radiative cooling time is much shorter than the 480 Myr characteristic age of MSP J0218+4232, the Peanut UHE $\gamma$-ray emission originates from recently injected leptons from this MSP. Accordingly, the cooling time is taken as the effective injection time $t$.

Using a distance of 3.15 kpc, the length and width of the Peanut strip correspond to physical scales of $r_\mathrm{d,\parallel} \simeq 250\ {\rm pc}$ and $r_\mathrm{d,\perp} \simeq 12.5\ {\rm pc}$, respectively. These dimensions constrain the diffusion coefficients along the two principal axes. The faster diffusion coefficient, aligned with the strip's major axis, is estimated to be $D_{\parallel} \simeq 3.3\times 10^{29}\times(E_\mathrm{e}/400\ {\rm TeV})^{\delta}\ {\rm cm}^2\ {\rm s}^{-1}$, where $\delta$ parameterizes the energy dependence. The insignificant detection of morphological variation  with energy suggests $\delta$ is close to 0.5. Adopting $\delta = 0.5$, we obtain $D_{\parallel,100} \simeq 1.7 \times 10^{29}\ {\rm cm}^2\ {\rm s}^{-1}$ at 100 TeV. This value is approximately 100 times larger than the diffusion coefficient inferred from the extended TeV halo around the Geminga pulsar \cite{2017Sci...358..911A} but remains compatible with estimates derived from Galactic cosmic-ray secondary measurements \cite{Guo:2015csa}. In the transverse direction, the perpendicular diffusion coefficient, $D_{\perp}=D_{\parallel}\times(r_\mathrm{d,\perp}/r_\mathrm{d,\parallel})^2\simeq D_{\parallel}/400$, is consistent with values estimated from the spatial extension of the Geminga TeV halo \cite{2017Sci...358..911A}. 

The above discussions adopt a uniform spectral shape across the entire region, as current statistics fail to reveal obvious spatial spectral variations. Nevertheless, we note that spectral gradients would be a key diagnostic of particle transport and radiative aging in such an extended system. The existing data do not yet have the discriminating power to robustly test for such gradients, and the assumption of a spatially uniform spectrum should be regarded as a working hypothesis limited by the current sensitivity.

\subsection{Other possible hidden accelerators}
Despite the MSP being a very likely leptonic accelerator for the Peanut’s $\gamma$-ray emission, the possibility of emission originating from other hidden leptonic or hadronic accelerators remains, e.g., an off-beam pulsar, an undetected microquasar, or even an object belonging to an entirely new PeVatron population. The detection of misaligned pulsars has been predicted and expected since the identification of the TeV halos around Geminga and Monogem~\cite{Fang:2023axu,HAWC:2024scl,HAWC:2017kbo}. This observed Peanut complex region potentially suggests the presence of a putative, non-aligned, middle-aged pulsar near the center. This hypothesis provides a natural explanation for the location of the highest-energy photon~(760 TeV) close to the center, which suggests a powerful particle acceleration and injection site at that position. Additionally, Jet-ISM interactions could also produce such asymmetric UHE $\gamma$-ray sources, as evidenced by microquasars like V4641 Sgr, in which asymmetric UHE $\gamma$ rays have been significantly detected~\cite{LHAASO:2024psv,Alfaro:2024cjd}. The highest-energy photons observed from V4641 Sgr exhibit comparable energies to those detected in the Peanut structure. Finally, the morphology of the Peanut is distinct from all previously known high-energy $\gamma$-ray objects, and UHE $\gamma$ rays serve as a unique probe for investigating new non-thermal astronomical phenomena. It is also possible that the Peanut is powered by an object belonging to an entirely new PeVatron population.

\section{Summary}
\label{sec:conclu}
Based on the unprecedented sensitivity of four years of survey data from LHAASO, this paper reports the discovery of a UHE $\gamma$-ray source with a distinctive peanut-like morphology. The source reaches a significance above 16$\sigma$ at energies $>$100 TeV and exhibits a spatial structure unlike any previously known high energy  $\gamma$-ray object, representing the emergence of a new class of $\gamma$-ray source and, correspondingly, a new type of PeVatron. The radiation spectrum from 10 to 760 TeV is well described by a power law with an exponential cutoff, characterized by a hard spectral index of $-$1.95, significantly harder than most other UHE $\gamma$-ray sources—and a cutoff energy of 250 TeV, which may reflect the maximum acceleration capability of the underlying source. At the brighter end of the source, a luminous MSP J0218+4232 shows a spatial coincidence with a significance of 3.4$\sigma$, suggesting a plausible physical association. The observed emission likely originates from electrons accelerated in the pulsar wind nebula of this MSP. After escaping the acceleration region, these electrons propagate anisotropically along the direction of the Galactic magnetic field and upscatter CMB photons via inverse Compton scattering, giving rise to the observed morphology. This positional coincidence provides the likely first observational evidence for a MSP wind nebula as the UHE $\gamma$-ray source. We cannot, however, entirely rule out the possibility that other unknown astrophysical objects dominate the emission from this region. Given the speculative nature of the peanut-shaped $\gamma$-ray emitter, future multi-wavelength observations, especially deep molecular line surveys and high-resolution X-ray imaging, are needed to distinguish between the possibilities. Therefore, we encourage deep follow-up observations with wide-field, multi-wavelength instruments to unveil the underlying physical processes of this peculiar source.

\section*{Conflict of interest}

The authors declare that they have no conflict of interest.

\section*{Acknowledgments}
We would like to thank the staff members of the National Facility for the Large High Altitude Air Shower Observatory (LHAASO) (https://cstr.cn/31117.02.LHAASO) who work at the LHAASO site above 4400~m above sea level year-round to maintain the detector and keep the water recycling system, power supply, and other experimental components operating smoothly. We are grateful to the Chengdu Management Committee of Tianfu New Area for their constant financial support for research using LHAASO data. We also appreciate the computing and data service support provided by the National High Energy Physics Data Center. This work was supported by the National Natural Science Foundation of China (12393851, 12261160362, 12393852, 12393853, 12393854, 12022502, 12205314, 12105301, 12105292, 12105294, 12005246, 12173039, and 12375108), the Department of Science and Technology of Sichuan Province, China (2024NSFJQ0060), the Project for Young Scientists in Basic Research of the Chinese Academy of Sciences (YSBR-061), and in Thailand by the National Science and Technology Development Agency (NSTDA) and the National Research Council of Thailand (NRCT): High-Potential Research Team Grant Program (N42A650868).\\

\section*{Author Contributions}
Zhe Li, as the project convener, led the overall study and drafted the manuscript. Shaoqiang Xi led the spatial morphology and spectral analyses, and contributed to the multi-wavelength analysis. Songzhan Chen contributed to the source parameter estimation, UHE $\gamma$-ray event selection, and manuscript revision. Dmitry Khangulyan provided key insights into the MSP interpretation and assisted in polishing the manuscript. Oscar Macias and Siming Liu actively contributed to discussions on the MSP interpretation and electron diffusion theories. The LHAASO Collaboration provided the essential data, infrastructure, and tools, with contributions spanning instrument design, construction, and operation, as well as software development and data calibration. All authors reviewed, discussed, and commented on the manuscript and the final results.\\


\section*{LHAASO Collaboration}
Zhen Cao$^{1,2,3}$,
F. Aharonian$^{3,4,5,6}$,
Y.X. Bai$^{1,3}$,
Y.W. Bao$^{7}$,
D. Bastieri$^{8}$,
X.J. Bi$^{1,2,3}$,
Y.J. Bi$^{1,3}$,
W. Bian$^{7}$,
A.V. Butkevich$^{9}$,
C.M. Cai$^{10}$
% ...
(The LHAASO Collaboration) \\ \\
$^{1}$ State Key Laboratory of Particle Astrophysics \& Experimental Physics Division \& Computing Center, Institute of High Energy Physics, Chinese Academy of Sciences, Beijing 100049, China\\
$^{2}$ University of Chinese Academy of Sciences, Beijing 100049, China\\
$^{3}$ TIANFU Cosmic Ray Research Center, Chengdu 610213,  China\\
$^{4}$ University of Science and Technology of China, Hefei 230026, China\\
$^{5}$ Yerevan State University, Yerevan 0025, Armenia\\
$^{6}$ Max-Planck-Institut for Nuclear Physics, Heidelberg 69029, Germany\\
$^{7}$ Tsung-Dao Lee Institute \& School of Physics and Astronomy, Shanghai Jiao Tong University, Shanghai 200240, China\\
$^{8}$ Center for Astrophysics, Guangzhou University, Guangzhou 510006, China\\
$^{9}$ Institute for Nuclear Research of Russian Academy of Sciences, Moscow 117312, Russia\\
$^{10}$ School of Physical Science and Technology \&  School of Information Science and Technology, Southwest Jiaotong University, Chengdu 610031, China\\
% ...

\end{document}